\documentclass[%
reprint,
superscriptaddress,
amsmath,amssymb,
aps,
floatfix,
prl
]{revtex4-2}

\usepackage{graphicx}
\usepackage{dcolumn}
\usepackage{bm}
\usepackage{hyperref}

\begin{document}

\newcommand{\alum}{$^{22}\mathrm{Al}$}
\newcommand{\magn}{$^{22}\mathrm{Mg}$}
\newcommand{\flor}{$^{22}\mathrm{F}$}
\newcommand{\neon}{$^{18}\mathrm{Ne}$}

\preprint{APS/123-QED}

\title{\textbf{A ground state $^{22}$Al halo is unlikely}}

\author{E. A. M. Jensen}
 \email[Contact author: ]{erik.jensen@chalmers.se}
 \affiliation{
 Institut for Fysik \& Astronomi, Aarhus Universitet, Aarhus, 8000, Denmark
 }%
 \affiliation{
 Institutionen för Fysik, Chalmers Tekniska Högskola, 412 96, Göteborg, Sweden
 }

\author{J. S. Nielsen}
 \affiliation{
 Institut for Fysik \& Astronomi, Aarhus Universitet, Aarhus, 8000, Denmark
 }

\author{B. S. O. Johansson}
 \affiliation{
 Institut for Fysik \& Astronomi, Aarhus Universitet, Aarhus, 8000, Denmark
 }

\author{A. Adams}
 \affiliation{
 Department of Physics and Astronomy, Michigan State University, East Lansing, Michigan 48824, USA
 }%
 \affiliation{
 Facility for Rare Isotope Beams, Michigan State University, East Lansing, Michigan 48824, USA
 }

\author{J. Dopfer}
 \affiliation{
 Department of Physics and Astronomy, Michigan State University, East Lansing, Michigan 48824, USA
 }%
 \affiliation{
 Facility for Rare Isotope Beams, Michigan State University, East Lansing, Michigan 48824, USA
 }

\author{C. S. Sumithrarachchi}
 \affiliation{
 Facility for Rare Isotope Beams, Michigan State University, East Lansing, Michigan 48824, USA
 }

\author{L. J. Sun}
 \affiliation{
 Facility for Rare Isotope Beams, Michigan State University, East Lansing, Michigan 48824, USA
 }

\author{L. E. Weghorn}
 \affiliation{
 Department of Physics and Astronomy, Michigan State University, East Lansing, Michigan 48824, USA
 }%
 \affiliation{
 Facility for Rare Isotope Beams, Michigan State University, East Lansing, Michigan 48824, USA
 }

\author{T. Wheeler}
 \affiliation{
 Department of Physics and Astronomy, Michigan State University, East Lansing, Michigan 48824, USA
 }%
 \affiliation{
 Facility for Rare Isotope Beams, Michigan State University, East Lansing, Michigan 48824, USA
 }%

\author{C. Wrede}
 \affiliation{
 Department of Physics and Astronomy, Michigan State University, East Lansing, Michigan 48824, USA
 }%
 \affiliation{
 Facility for Rare Isotope Beams, Michigan State University, East Lansing, Michigan 48824, USA
 }

\author{M. J. G. Borge}
 \affiliation{
 Instituto de Estructura de la Materia, CSIC, Madrid, 28006, Spain
 }

\author{O. Tengblad}
 \affiliation{
 Instituto de Estructura de la Materia, CSIC, Madrid, 28006, Spain
 }

\author{M. Madurga}
 \affiliation{
 Department of Physics and Astronomy, University of Tennessee, Knoxville, Tennessee 37996, USA
 }

\author{B. Jonson}
 \affiliation{
 Institutionen för Fysik, Chalmers Tekniska Högskola, 412 96, Göteborg, Sweden
 }

\author{K. Riisager}
 \affiliation{
 Institut for Fysik \& Astronomi, Aarhus Universitet, Aarhus, 8000, Denmark
 }

\author{H. O. U. Fynbo}
 \affiliation{
 Institut for Fysik \& Astronomi, Aarhus Universitet, Aarhus, 8000, Denmark
 }

\date{\today}

\begin{abstract}%
We report the decisive resolution of the ground state spin and parity of the proton-dripline nucleus $^{22}$Al, a prime candidate for a proton halo.
The resolution stems from the first $\beta$-delayed charged particle emission experiment in the Gas Stopping Area at the Facility for Rare Isotope Beams (FRIB), leveraging high-intensity, low-energy beams extracted from the Advanced Cryogenic Gas Stopper (ACGS).
The pristine beam quality from FRIB and the ACGS enabled a sensitive particle identification technique using thin silicon detectors, allowing for the suppression of the dominant proton background and the first observation of the weak $\beta$-delayed $\alpha$ transition from the Isobaric Analog State in $^{22}$Mg to the $^{18}$Ne ground state.
This observation uniquely fixes the $^{22}$Al ground state as $4^+$.
The valence proton is confined by a dominant $d$-wave centrifugal barrier which, combined with the Coulomb repulsion, hinders the tunneling required for halo formation despite the exceptionally low proton separation energy of $^{22}$Al.
\end{abstract}

\maketitle



Near the limits of nuclear stability, the classical picture of the nucleus as a compact liquid drop breaks down, giving rise to the nuclear ``halo''.
This exotic structure emerges in loosely bound systems where a vanishing separation energy allows valence nucleons to tunnel far into the classically forbidden region.
The resulting spatially extended wavefunction creates a unique state of dilute nuclear matter that defies standard geometric scaling laws, serving as a distinct testbed for quantum phenomena at the edge of unbinding.

While weak binding is a necessary condition for halo formation, it is not sufficient; the quantum tunneling required to sustain a halo is critically sensitive to the confining potential barriers \cite{Rii94,Jen04}.
For a halo to develop, the valence nucleon must not be inhibited by a high potential barrier.
Consequently, the orbital angular momentum, $l$, plays a decisive role.
The centrifugal barrier, proportional to $l(l+1)$, strongly suppresses the tunneling of nucleons in high-$l$ orbitals, effectively confining the wavefunction to the nuclear core.
Therefore, halos are predominantly associated with $s$-waves ($l=0$) or occasionally $p$-waves ($l=1$).
This structural fragility is further compounded in proton-rich nuclei by the Coulomb barrier.
Unlike neutrons, which can form halos relatively easily in light nuclei (such as $^{11}\text{Li}$ or $^{11}\text{Be}$), loosely bound protons are contained by the repulsive long-range Coulomb potential, which tends to truncate the outer tail of the wavefunction.
As a result, genuine proton halos are rare and remain a subject of intense debate \cite{Tan96,Jon04,Leh22,Zha24,Zha25}.

Proton halos in the $sd$-shell have been envisaged since \cite{Bro96}, often in light isotopes of Si and P.
Recently, much interest has been given to the lightest bound Al isotope, $^{22}\mathrm{Al}$.
The recent interest started with the observation \cite{Lee20} of an asymmetry between the $\beta$-decays into the lowest excited $1^+$ states in $^{22}\mathrm{Al}$ and its mirror nucleus $^{22}\mathrm{F}$ that was attributed to a proton halo structure.
Two detailed theory papers \cite{Zha24,Pap25} do not support this conclusion, in particular not for the ground state, but acknowledge the incomplete experimental situation that has been partially alleviated by recent mass measurements \cite{Cam24,Sun24} which have established the proton separation energy of $^{22}\mathrm{Al}$ to be $S_p = 100.3(8)$ keV
\footnote{The proton separation energy $S_p = 100.3(8)$ keV of $^{22}\mathrm{Al}$ is the weighted average of 100.4(8) keV from \cite{Cam24} and 90(10) keV from \cite{Sun24}.}.

The realization of a halo structure depends entirely on the quantum numbers of the valence nucleon, which dictate the height of the centrifugal barrier.
The structure of $^{22}\text{Al}$ can be modeled as a valence proton coupled to a $^{21}\text{Mg}$($5/2$) core, but the spin and parity, $J^\pi$, of the $^{22}\text{Al}$ ground state have remained ambiguous, with experimental evidence and theoretical models oscillating between a $3^+$ and a $4^+$ assignment, as discussed below.
This distinction is the primary determining factor for the existence of a halo.
A $J^\pi = 3^+$ assignment may allow the valence proton to occupy an $s_{1/2}$ orbital ($l=0$) relative to the core, facing no centrifugal barrier and allowing for the possible formation of an extended proton halo.
Conversely, a $J^\pi = 4^+$ assignment places the proton in a $d_{5/2}$ orbital ($l=2$), where the combined centrifugal and Coulomb barriers would confine the proton, resulting in a more standard nuclear structure.


The structural interpretation of \alum{} is informed by knowledge from its mirror nucleus, \flor{}, which decays to the stable and well-characterized \cite{Bas15} nuclide $^{22}$Ne.
Together with \magn{} and $^{22}$Na, these nuclides constitute an isospin $T=2$ quintet.
While the $\beta$-decay strength from \flor{} to excited states in $^{22}$Ne restricts the \flor{} ground state to an even parity and a spin of $J=3$ or $4$ \cite{Dav74}, a precise assignment remains elusive.
Conflicting interpretations arise from reaction studies populating the analog 5.523 MeV state in $^{22}$Ne:
$^{18}$O($^{6}$Li,d) and $^{20}$Ne(t,p) favor $J=4$ \cite{Ana77,Mao94}, while $^{13}$C($^{11}$B,d) favors $J=3$ \cite{Sza83}.

Under strict isospin symmetry, the ground state of \alum{} shares the tentative $J^{\pi}=(3,4)^+$ assignment of \flor{}.
However, the level structure is complicated by the presence of a first excited state in \flor{} at just 72 keV.
The de-excitation of this state to the ground state is found in \cite{Lee07} to change the spin by one unit and is assigned $3^+$, assuming the ground state of \flor{} is $4^+$.
What is clear, based on the available literature on the $T=2$ quintet, is that the two lowest-lying states in \flor{} are $3^+$ and $4^+$; two $3^+$ states in such close proximity is highly unlikely.
It is, however, unclear which of the two states has the lowest energy in \flor{}.
In the proton-rich \alum{}, the Thomas-Ehrman shift—driven by the spatial extension of the unbound proton wavefunction—could lower the energy of the analog $3^+$ state, which is associated with a low angular momentum $l$.
This effect could conceivably induce an inversion in the ground and first excited states in \alum{} relative to \flor{}, if the first excited state in \flor{} is indeed a $3^+$ state.
In other words, the ground state of \alum{} could be a $3^+$ state regardless of the ground state of \flor{} possibly being a $4^+$ state.


We report the first observation of $\beta$-delayed $\alpha$ emission from the \alum{} ground state, proceeding via the Isobaric Analog State (IAS) in \magn{} to the ground state of \neon{}.
As detailed below, this observation (1) uniquely determines the spin and parity of the \alum{} ground state to be $4^+$, (2) resolves the long-standing ambiguity in the $A=22$ isospin quintet and (3) challenges the status of \alum{} as a halo nucleus.

\begin{figure}
\includegraphics[width=0.98\columnwidth]{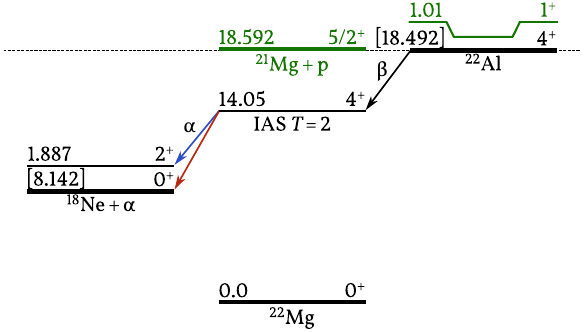}
\caption{\label{fig:decay-scheme}%
Decay scheme for the $\beta$-delayed $\alpha$ emission of $^{22}$Al.
The newly observed transition to the $^{18}$Ne ground state is highlighted in red.
The mass and proton separation energy of \alum{} are adopted from \cite{Cam24}, and the IAS energy is derived from our measured $\beta$-delayed proton spectrum.
Other level parameters are adopted from \cite{Til95,Lee20,Fir15}.
The 1.01 MeV $1^+$ state in $^{22}$Al is the lowest-lying known excited state, and is particle-unbound.
Its energy is recalculated, based on the new mass measurement in \cite{Cam24}, increasing it from 0.91 MeV reported in \cite{Lee20}.
}
\end{figure}

The relevant decay scheme is presented in Fig. \ref{fig:decay-scheme}.
Previous studies at GANIL \cite{Bla97,Ach06} and HIRFL \cite{Wu21} identified the $\beta$-delayed $\alpha$ branch to the first excited ($2^+$) state of \neon{} (indicated by the blue arrow).
However, the critical transition to the $0^+$ ground state (indicated by the red arrow) remained unobserved in these experiments, likely due to the overwhelming background from $\beta$-delayed protons.

The present measurement was performed at the Facility for Rare Isotope Beams (FRIB).
A 5 kW primary beam of $^{36}$Ar was accelerated to 210 MeV/u before impinging on a 8.07 mm carbon production target.
The resulting in-flight cocktail beam was momentum-to-charge-separated in the Advanced Rare Isotope Separator (ARIS) \cite{Por23,Fuk23}, reducing the beam energy to 106 MeV/u.
Subsequently, the beam was guided to the Gas Stopping Area for further momentum compression, thermalization and purification.
A pure, low-energy 30 keV beam of \alum{} was extracted from the Advanced Cryogenic Gas Stopper (ACGS) \cite{Lun20,Rin21} and implanted into a thin carbon foil surrounded by a compact array of silicon detector telescopes.
Just outside of the vacuum chamber, containing the silicon detector telescopes, were two High-Purity Germanium detectors from the LIBRA setup \cite{Sun25}, flanking the chamber.
The low beam energy and the use of thin (60--70 µm) $\Delta E$ detectors allowed the dominant high-energy protons above 2--3 MeV to punch through the thin detectors, depositing only a fraction of their energy, while fully stopping $\alpha$ particles up to $\sim$9 MeV \cite{Zie10}.
This suppression of the proton background, crucial for isolating the rare $\alpha$ decay channels, was made possible by the characteristics of the low energy beams extracted from the ACGS.

\begin{figure}
\includegraphics[width=0.98\columnwidth]{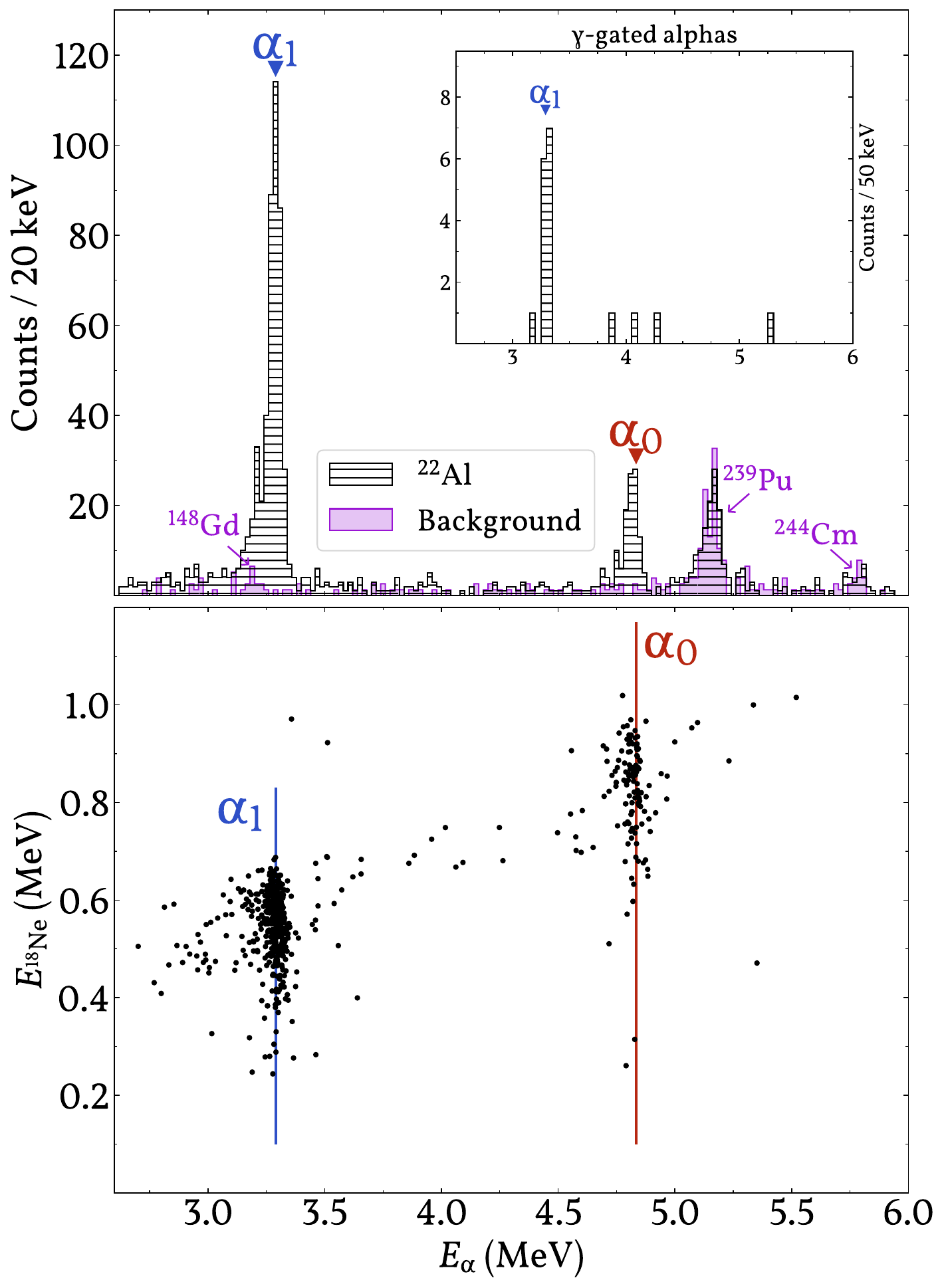}
\caption{\label{fig:spectra}%
\textbf{Top:} Singles spectrum of $\alpha$-particle kinetic energies, $E_\alpha$, extracted from the $\Delta E$ detectors.
The labels $\alpha_0$ and $\alpha_1$ denote transitions from the IAS in \magn{} to the $0^+$ ground and $2^+$ first excited states of \neon{}, respectively.
The background spectrum (scaled from 25 h to the 33 h \alum{} measurement time) is dominated by pre-implanted calibration sources.
The inset shows coincidences with the 1.887 MeV $\gamma$-transition in \neon{}.
\textbf{Bottom:} \neon{} recoil energies, $E_{^{18}\mathrm{Ne}}$, in coincidence with $\alpha$ particles in opposing silicon detectors.
The $\alpha$ particles are found above the proton punch through thresholds in the $\Delta E$ detectors (as in the singles spectrum), and any signal in opposing detectors is considered a recoil.
}
\end{figure}

The upper panel of Fig. \ref{fig:spectra} displays the energy spectrum of $\alpha$ particles stopped in the $\Delta E$ detectors.
Two distinct peaks associated with the decay of \alum{} are visible above the background.
The background arises from long-lived calibration sources implanted prior to the experiment.
The peak at 3.29 MeV ($\alpha_1$) corresponds to the previously known transition to the $^{18}$Ne($2^+$) state.
The newly observed peak at 4.83 MeV ($\alpha_0$) corresponds to the transition to the $^{18}$Ne($0^+$) ground state.

The identification is supported by coincidence measurements.
The lower panel of Fig. \ref{fig:spectra} shows $^{18}$Ne-$\alpha$ coincidences demonstrating the detection of low-energy signals from the opposing silicon detectors, corresponding to $^{18}$Ne nuclei recoiling with the expected kinetic energies of 1.07 MeV and 0.73 MeV for the ground and excited state transitions, respectively.
We emphasize that the pure, low-energy beam extracted from the ACGS enabled the measurement of the low-energy \neon{} recoils.
Furthermore, a gate on the 1.887 MeV $\gamma$-ray line (the $2^+ \to 0^+$ transition in \neon{}) isolates the $\alpha_1$ branch, as shown in the inset of the upper panel.
The particle energies in Fig. \ref{fig:spectra} have been corrected for energy losses specific to $\alpha$ particles in the detector dead layers and for the pulse height defect (PHD) inherent to silicon detectors calibrated with protons.
The observed broadening of the recoil energies is consistent with the broader distribution of energy losses of the heavy ions in the dead layers and with PHD uncertainties.

\begin{table}
\caption{\label{tab:br-penet}%
Kinetic energies, $E_\alpha$, relative branching ratios, $BR$, and calculated single-particle penetrabilities, $P_l$, for $\alpha$ emission from the \magn{} IAS with orbital angular momentum $l$.
Systematic energy uncertainties ($\sim$30 keV) dominate over statistical errors. $P_l$ values are median results from a uniform sampling of channel radii $R = r_0(A^{1/3}_1 + A^{1/3}_2) \, \mathrm{fm}$ with $r_0 \in [1.2, 1.4]$ and $A_1=18, A_2=4$.
}
\begin{ruledtabular}
\begin{tabular}{ r | c c }
                 & $\alpha_0 \, (l=4) \rightarrow \,^{18}\textrm{Ne}(0^+)$   & $\alpha_1 \, (l=2) \rightarrow \,^{18}\textrm{Ne}(2^+)$ \\
\colrule
$E_\alpha$ (MeV) & 4.83                                                      & 3.29                                                    \\
$BR$ (\%)        & 22(2)                                                     & 78(2)                                                   \\
$P_l$            & 0.44(9)                                                   & 0.33(6)
\end{tabular}
\end{ruledtabular}
\end{table}

The relative branching ratios are summarized in Table \ref{tab:br-penet}. The transition to the $2^+$ state is favored, 78\%, over the ground state, 22\%, despite the latter having a slightly larger barrier penetrability.
The ratio of the observed branching ratios, ${BR}_{\alpha_0}/{BR}_{\alpha_1} \approx 0.28$, contrasts with the ratio of penetrabilities, $P_{\alpha_0}/P_{\alpha_1} \approx 1.3$, indicating a hindrance of the ground state transition by a factor of roughly 4--5.

The observation of the $\alpha_0$ branch allows for a definitive assignment of the \alum{} ground state spin and parity:
The emission of an $\alpha$ particle to a $0^+$ state requires, by conservation of angular momentum and parity, that the parent state has natural parity and $J^\pi = l^\pi$.
Since the IAS in \magn{} is populated via superallowed Fermi decay, it shares the spin and parity of the \alum{} ground state.
A $J^\pi = 3^+$ assignment would require $l=3$ for decay to a $0^+$ daughter, which implies negative parity, leading to a violation of parity conservation.
Therefore, the observation of the $\alpha_0$ branch mandates that the IAS—and consequently the ground state of \alum{}—is $4^+$.


The newly determined proton separation energy of \alum{}, $S_p = 100.3(8)$ keV \cite{Cam24,Sun24}, is exceptionally low and comparable to the archetype proton halo nucleus $^{8}$B \cite{Sme99,Kor18} with $S_p = 136(1)$ keV.
This proximity to the dripline has fueled speculation that \alum{} might exhibit a proton halo.
However, halo formation is not determined solely by binding energy; it requires a structural configuration that allows the valence nucleon to tunnel through the confining potentials.

If the ground state of \alum{} were $3^+$, the valence proton could occupy an $s_{1/2}$ ($l=0$) orbital coupled to the $^{21}$Mg($5/2^+$) core ground state.
Lacking a centrifugal barrier, such a configuration would favor the formation of an extended halo.
In contrast, the $4^+$ assignment mandates that a proton coupled to the $^{21}$Mg ground state must occupy a $d_{5/2}$ ($l=2$) orbital.
An $s$-wave component in the $4^+$ state is only possible via coupling to high-lying excited states of the core (e.g., $7/2^+, 9/2^+$), a configuration that is energetically suppressed.



This qualitative picture is supported by recent microscopic calculations.
Relativistic Hartree-Bogoliubov studies treating \alum{} as a deformed or triaxial system \cite{Zha24,Pap25} indicate that even under favorable conditions, the ground state wavefunction is dominated ($>90\%$) by the $d$-wave component.
Consequently, these models predict no significant enhancement of the root-mean-square radius relative to neighboring isotopes.
Furthermore, recent \emph{ab-initio} calculations \cite{Li23} suggest that sizable Thomas-Ehrman shifts are ubiquitous in $sd$-shell mirror nuclei; these shifts can rationalize the low separation energy and level ordering in the $A=22$ system without necessitating the spatial delocalization characteristic of a halo.

We therefore conclude that despite the vanishingly small separation energy, the $4^+$ ground state of \alum{} is likely a standard nuclear system confined by high potential barriers.
However, we note that a subtle enhancement of the surface density or a "soft" tail in the wavefunction cannot be strictly excluded by spectroscopy alone.
The ultimate confirmation of the compact nature of \alum{} awaits a direct measurement of its charge radius.


The successful delivery of exotic nuclei at low energy from the ACGS at FRIB represents a significant advance for experimental studies near the driplines.
In this work, the pure, low-energy beams from the ACGS enabled both a sensitive particle identification technique that separated rare low-energy $\alpha$ particles from the dominant proton background and the detection of coincident low-energy \neon{} recoils.
This capability was instrumental in the first observation of the $\beta$-delayed $\alpha$ decay to the ground state of \neon{}.

This observation has immediate and decisive consequences for the structure of \alum{}.
By unambiguously assigning a spin and parity of $4^+$ to the ground state, we have ruled out the $s$-wave coupling required for a pronounced proton halo.
Instead, the valence proton is confined by a $d$-wave centrifugal barrier which, in concert with the Coulomb barrier, renders the formation of a halo highly improbable despite the vanishingly small proton separation energy.

Furthermore, the assignment of $4^+$ to the \alum{} ground state has implications for the low-energy structure of \flor{} with its first excited state located just 72 keV above its ground state.
These two states are $3^+$ and $4^+$, but their ordering is unclear, as explained above.
Our result does not uniquely settle this, but strongly favors $4^+$ also for the \flor{} ground state.

The lowest-lying known excited $1^+$ state in \alum{} (Fig. \ref{fig:decay-scheme}) is unbound and thus cannot form a halo \cite{Zha24}.
Should an excited $3^+$ state in \alum{} lie below the proton separation energy $S_p = 100.3(8)$ keV, it remains a candidate for extended structure.

Finally, while our spectroscopic result provides the angular momentum constraint deemed critical in \cite{Cam24}, the ultimate quantification of the proton wavefunction's spatial extent requires a direct observable. We advocate for a charge radius measurement as the final arbiter.
While interaction cross-section measurements could provide complementary evidence, laser spectroscopy of low-energy beams—now feasible in the FRIB Gas Stopping Area—appears to be the most direct path to conclusively settle the question of the \alum{} radius.

\begin{acknowledgments}


This material is based upon work supported by the U.S. Department of Energy, Office of Science, Office of Nuclear Physics and used resources of the Facility for Rare Isotope Beams (FRIB) Operations, which is a DOE Office of Science User Facility under Award Number DE-SC0023633.

The authors acknowledge the support from the Independent Research Fund Denmark, Project Nos. 9040-00076B, 2032-00066B and 4283-00172B, from the Swedish Research Council, Project No. 2022-04248, from the U.S. National Science Foundation under Grant Nos. PHY-1913554, PHY-2209429, and PHY-2514797, from the Spanish MICIU/AEI/10.13039/501100011033, and from FEDER, EU, under Project No. PID2022-140162NB-I00.

\end{acknowledgments}

\bibliography{main}

@PREAMBLE{
 "\providecommand{\noopsort}[1]{}" 
 # "\providecommand{\singleletter}[1]{#1}%" 
}

@ARTICLE{Rii94,
   author       = "K. Riisager",
   title        = "Nuclear halo states",
   journal      = "Rev.\ Mod.\ Phys.",
   volume       = "66",
   pages        = "1105",
   year         = "1994",
   url          = "https://doi.org/10.1103/RevModPhys.66.1105"
}

@ARTICLE{Jen04,
   author       = "A. S. Jensen and K. Riisager and D. V. Fedorov and E. Garrido",
   title        = "Structure and reactions of quantum halos",
   journal      = "Rev.\ Mod.\ Phys.",
   volume       = "76",
   pages        = "215",
   year         = "2004",
   url          = "https://doi.org/10.1103/RevModPhys.76.215"
}

@ARTICLE{Bro96,
   author       = "B. A. Brown and P. G. Hansen",
   title        = "Proton halos in the $1s0d$ shell",
   journal      = "Phys.\ Lett.\ B",
   volume       = "381",
   pages        = "391",
   year         = "1996",
   url          = "https://doi.org/10.1016/0370-2693(96)00634-X"
}

@ARTICLE{Lee20,
   author       = "J. Lee and X. X. Xu and K. Kaneko and Y. Sun and C. J. Lin and L. J. Sun and P. F. Liang and Z. H. Li and J. Li and H. Y. Wu and D. Q. Fang and J. S. Wang and Y. Y. Yang and C. X. Yuan and Y. H. Lam and Y. T. Wang and K. Wang and J. G. Wang and J. B. Ma and J. J. Liu and P. J. Li and Q. Q. Zhao and L. Yang and N. R. Ma and D. X. Wang and F. P. Zhong and S. H. Zhong and F. Yang and H. M. Jia and P. W. Wen and M. Pan and H. L. Zang and X. Wang and C. G. Wu and D. W. Luo and H. W. Wang and C. Li and C. Z. Shi and M. W. Nie and X. F. Li and H. Li and P. Ma and Q. Hu and G. Z. Shi and S. L. Jin and M. R. Huang and Z. Bai and Y. J. Zhou and W. H. Ma and F. F. Duan and S. Y. Jin and Q. R. Gao and X. H. Zhou and Z. G. Hu and M. Wang and M. L. Liu and R. F. Chen and X. W. Ma",
   title        = "Large {I}sospin {A}symmetry in $^{22}\mathrm{Si}$/$^{22}\mathrm{O}$ {M}irror {G}amow-{T}eller {T}ransitions {R}eveals the {H}alo {S}tructure of $^{22}\mathrm{Al}$",
   journal      = "Phys.\ Rev.\ Lett.",
   volume       = "125",
   pages        = "192503",
   year         = "2020",
   url          = "https://doi.org/10.1103/PhysRevLett.125.192503"
}

@ARTICLE{Zha24,
   author       = "K. Y. Zhang and C. Pan and S. Wang",
   title        = "Examination of the evidence for a proton halo in $^{22}\mathrm{Al}$",
   journal      = "Phys.\ Rev.\ C",
   volume       = "110",
   pages        = "014320",
   year         = "2024",
   url          = "https://doi.org/10.1103/PhysRevC.110.014320"
}

@ARTICLE{Pap25,
   author       = "P. Papakonstantinou and M. Mun and C. Pan and K. Zhang",
   title        = "Proton halo structures and $^{22}\mathrm{Al}$",
   journal      = "Phys.\ Rev.\ C",
   volume       = "112",
   pages        = "044301",
   year         = "2025",
   url          = "https://doi.org/10.1103/yt6n-bp9f"
}

@ARTICLE{Cam24,
   author       = "S. E. Campbell and G. Bollen and B. A. Brown and A. Dockery and C. M. Ireland and K. Minamisono and D. Puentes and B. J. Rickey and R. Ringle and I. T. Yandow and K. Fossez and A. Ortiz-Cortes and S. Schwarz and C. S. Sumithrarachchi and A. C. C. Villari",
   title        = "Precision {M}ass {M}easurement of the {P}roton {D}ripline {H}alo {C}andidate $^{22}\mathrm{Al}$",
   journal      = "Phys.\ Rev.\ Lett.",
   volume       = "132",
   pages        = "152501",
   year         = "2024",
   url          = "https://doi.org/10.1103/PhysRevLett.132.152501"
}

@ARTICLE{Bas15,
   author       = "M. S. Basunia",
   title        = "Nuclear {D}ata {S}heets for {A} = 22",
   journal      = "Nucl.\ Data\ Sheets",
   volume       = "127",
   pages        = "69",
   year         = "2015",
   url          = "https://doi.org/10.1016/j.nds.2015.07.002"
}

@ARTICLE{Dav74,
   author       = "C. N. Davids and D. R. Goosman and D. E. Alburger and A. Gallmann and G. Guillaume and D. H. Wilkinson and W. A. Lanford",
   title        = "$\beta$ decay of $^{22}\mathrm{F}$",
   journal      = "Phys.\ Rev.\ C",
   volume       = "9",
   pages        = "216",
   year         = "1974",
   url          = "https://doi.org/10.1103/PhysRevC.9.216"
}

@ARTICLE{Ana77,
   author       = "N. Anantaraman and H. E. Gove and J. P. Trentelman and J. P. Draayer and F. C. Jundt",
   title        = "A study of the $^{18}\mathrm{O}$($^{6}\mathrm{Li}$,d)$^{22}\mathrm{Ne}$ reaction at 32 {MeV}",
   journal      = "Nucl.\ Phys.\ A",
   volume       = "276",
   pages        = "119",
   year         = "1977",
   url          = "https://doi.org/10.1016/0375-9474(77)90162-2"
}

@ARTICLE{Mao94,
   author       = "Z. Q. Mao and H. T. Fortune",
   title        = "Mechanism of $^{20}\mathrm{Ne}$(t,p) and nuclear structure of $^{22}\mathrm{Ne}$",
   journal      = "Phys.\ Rev.\ C",
   volume       = "50",
   pages        = "2116",
   year         = "1994",
   url          = "https://doi.org/10.1103/PhysRevC.50.2116"
}

@ARTICLE{Sza83,
   author       = "E. M. Szanto and A. {Szanto de Toledo} and H. V. Klapdor and G. Rosner and M. Schrader",
   title        = "Yrast and high-spin states in $^{22}\mathrm{Ne}$",
   journal      = "Nucl.\ Phys.\ A",
   volume       = "404",
   pages        = "142",
   year         = "1983",
   url          = "https://doi.org/10.1016/0375-9474(83)90419-0"
}

@ARTICLE{Lee07,
   author       = "S. Lee and S. L. Tabor and A. Volya and A. Aguilar and P. C. Bender and T. A. Hinners and C. R. Hoffman and M. Perry and V. Tripathi",
   title        = "Electromagnetic transitions in neutron-rich $^{22}\mathrm{F}$",
   journal      = "Phys.\ Rev.\ C",
   volume       = "76",
   pages        = "034308",
   year         = "2007",
   url          = "https://doi.org/10.1103/PhysRevC.76.034308"
}

@ARTICLE{Til95,
   author       = "D. R. Tilley and H. R. Weller and C. M. Cheves and R. M. Chasteler",
   title        = "Energy levels of light nuclei {A} = 18–19",
   journal      = "Nucl.\ Phys.\ A",
   volume       = "595",
   pages        = "1",
   year         = "1995",
   url          = "https://doi.org/10.1016/0375-9474(95)00338-1"
}

@ARTICLE{Bla97,
   author       = "B. Blank and F. Boué and S. Andriamonje and S. Czajkowski and R. {Del Moral} and J. P. Dufour and A. Fleury and P. Pourre and M. S. Pravikoff and N. A. Orr and K.-H. Schmidt and E. Hanelt",
   title        = "The spectroscopy of $^{22}\mathrm{Al}$: a $\beta$p, $\beta$2p and $\beta \alpha$ emitter",
   journal      = "Nucl.\ Phys.\ A",
   volume       = "615",
   pages        = "52",
   year         = "1997",
   url          = "https://doi.org/10.1016/S0375-9474(96)00483-6"
}

@ARTICLE{Ach06,
   author       = "N. L. Achouri and F. {de Oliveira Santos} and M. Lewitowicz and B. Blank and J. Äystö and G. Canchel and S. Czajkowski and P. Dendooven and A. Emsallem and J. Giovinazzo and N. Guillet and A. Jokinen and A. M. Laird and C. Longour and K. Peräjärvi and N. Smirnova and M. Stanoiu and J.-C. Thomas",
   title        = "The $\beta$-decay of $^{22}\mathrm{Al}$",
   journal      = "Eur.\ Phys.\ J.\ A",
   volume       = "27",
   pages        = "287",
   year         = "2006",
   url          = "https://doi.org/10.1140/epja/i2005-10274-0"
}

@ARTICLE{Wu21,
   author       = "C. G. Wu and H. Y. Wu and J. G. Li and D. W. Luo and Z. H. Li and H. Hua and X. X. Xu and C. J. Lin and J. Lee and L. J. Sun and P. F. Liang and C. X. Yuan and Y. Y. Yang and J. S. Wang and D. X. Wang and F. F. Duan and Y. H. Lam and P. Ma and Z. H. Gao and Q. Hu and Z. Bai and J. B. Ma and J. G. Wang and F. P. Zhong and Y. Jiang and Y. Liu and D. S. Hou and R. Li and N. R. Ma and W. H. Ma and G. Z. Shi and G. M. Yu and D. Patel and S. Y. Jin and Y. F. Wang and Y. C. Yu and Q. W. Zhou and P. Wang and L. Y. Hu and S. Q. Fan and X. Wang and H. L. Zang and P. J. Li and Q. Q. Zhao and L. Yang and P. W. Wen and F. Yang and H. M. Jia and G. L. Zhang and M. Pan and X. Y. Wang and H. H. Sun and Z. G. Hu and M. L. Liu and R. F. Chen and W. Q. Yang and S. Q. Hou and J. J. He and Y. M. Zhao and F. R. Xu and H. Q. Zhang",
   title        = "$\beta$-decay spectroscopy of the proton drip-line nucleus $^{22}\mathrm{Al}$",
   journal      = "Phys.\ Rev.\ C",
   volume       = "104",
   pages        = "044311",
   year         = "2021",
   url          = "https://doi.org/10.1103/PhysRevC.104.044311"
}

@ARTICLE{Lun20,
   author       = "K. R. Lund and G. Bollen and D. Lawton and D. J. Morrissey and J. Ottarson and R. Ringle and S. Schwarz and C. S. Sumithrarachchi and A. C. C. Villari and J. Yurkon",
   title        = "Online tests of the {A}dvanced {C}ryogenic {G}as {S}topper at {NSCL}",
   journal      = "Nucl.\ Instrum.\ Methods\ Phys.\ Res.\ B",
   volume       = "463",
   pages        = "378",
   year         = "2020",
   url          = "https://doi.org/10.1016/j.nimb.2019.04.053"
}

@ARTICLE{Rin21,
   author       = "R. Ringle and G. Bollen and K. Lund and C. Nicoloff and S. Schwarz and C. S. Sumithrarachchi and A. C. C. Villari",
   title        = "Particle-in-cell techniques for the study of space charge effects in the {A}dvanced {C}ryogenic {G}as {S}topper",
   journal      = "Nucl.\ Instrum.\ Methods\ Phys.\ Res.\ B",
   volume       = "496",
   pages        = "61",
   year         = "2021",
   url          = "https://doi.org/10.1016/j.nimb.2021.03.020"
}

@ARTICLE{Zie10,
   author       = "J. F. Ziegler and M. D. Ziegler and J. P. Biersack",
   title        = "{SRIM} – The stopping and range of ions in matter (2010)",
   journal      = "Nucl.\ Instrum.\ Methods\ Phys.\ Res.\ B",
   volume       = "268",
   pages        = "1818",
   year         = "2010",
   url          = "https://doi.org/10.1016/j.nimb.2010.02.091"
}

@ARTICLE{Kor18,
   author       = "G. A. Korolev and A. V. Dobrovolsky and A. G. Inglessi and G. D. Alkhazov and P. Egelhof and A. Estradé and I. Dillmann and F. Farinon and H. Geissel and S. Ilieva and Y. Ke and A. V. Khanzadeev and O. A. Kiselev and J. Kurcewicz and X. C. Le and Y. A. Litvinov and G. E. Petrov and A. Prochazka and C. Scheidenberger and L. O. Sergeev and H. Simon and M. Takechi and S. Tang and V. Volkov and A. A. Vorobyov and H. Weick and V. I. Yatsoura",
   title        = "Halo structure of $^{8}\mathrm{B}$ determined from intermediate energy proton elastic scattering in inverse kinematics",
   journal      = "Phys.\ Lett.\ B",
   volume       = "780",
   pages        = "200",
   year         = "2018",
   url          = "https://doi.org/10.1016/j.physletb.2018.03.013"
}

@ARTICLE{Li23,
   author       = "H. H. Li and Q. Yuan and J. G. Li and M. R. Xie and S. Zhang and Y. H. Zhang and X. X. Xu and N. Michel and F. R. Xu and W. Zuo",
   title        = "Investigation of isospin-symmetry breaking in mirror energy difference and nuclear mass with \emph{ab initio} calculations",
   journal      = "Phys.\ Rev.\ C",
   volume       = "107",
   pages        = "014302",
   year         = "2023",
   url          = "https://doi.org/10.1103/PhysRevC.107.014302"
}

@ARTICLE{Sun24,
   author       = "M. Z. Sun and Y. Yu and X. P. Wang and M. Wang and J. G. Li and Y. H. Zhang and K. Blaum and Z. Y. Chen and R. J. Chen and H. Y. Deng and C. Y. Fu and W. W. Ge and W. J. Huang and H. Y. Jiao and H. H. Li and H. F. Li and Y. F. Luo and T. Liao and Y. A. Litvinov and M. Si and P. Shuai and J. Y. Shi and Q. Wang and Y. M. Xing and X. Xu and H. S. Xu and F. R. Xu and Q. Yuan and T. Yamaguchi and X. L. Yan and J. C. Yang and Y. J. Yuan and X. H. Zhou and X. Zhou and M. Zhang and Q. Zeng",
   title        = "Ground-state mass of $^{22}\mathrm{Al}$ and test of state-of-the-art \emph{ab initio} calculations",
   journal      = "Chin.\ Phys.\ C",
   volume       = "48",
   pages        = "034002",
   year         = "2024",
   url          = "http://dx.doi.org/10.1088/1674-1137/ad1a0a"
}

@ARTICLE{Por23,
   author       = "M. Portillo and B. M. Sherrill and Y. Choi and M. Cortesi and K. Fukushima and M. Hausmann and E. Kwan and S. Lidia and P. N. Ostroumov and R. Ringle and M. K. Smith and M. Steiner and O. B. Tarasov and A. C. C. Villari and T. Zhang",
   title        = "{Commissioning of the Advanced Rare Isotope Separator ARIS at FRIB}",
   journal      = "Nucl.\ Instrum.\ Methods\ Phys.\ Res.\ B",
   volume       = "540",
   pages        = "151",
   year         = "2023",
   url          = "https://doi.org/10.1016/j.nimb.2023.04.025"
}

@ARTICLE{Fuk23,
   author       = "K. Fukushima and M. Cortesi and M. Hausmann and E. Kwan and P. N. Ostroumov and M. Portillo and B. M. Sherrill and M. Smith and M. Steiner and T. Zhang",
   title        = "{Simulation studies for beam commissioning at FRIB Advanced Rare Isotope Separator}",
   journal      = "Nucl.\ Instrum.\ Methods\ Phys.\ Res.\ B",
   volume       = "541",
   pages        = "53",
   year         = "2023",
   url          = "https://doi.org/10.1016/j.nimb.2023.04.038"
}

@ARTICLE{Fir15,
   author       = "R. B. Firestone",
   title        = "{Nuclear Data Sheets for A = 21}",
   journal      = "Nucl.\ Data\ Sheets",
   volume       = "127",
   pages        = "1",
   year         = "2015",
   url          = "https://doi.org/10.1016/j.nds.2015.07.001"
}

@ARTICLE{Leh22,
   author       = "C. Lehr and F. Wamers and F. Aksouh and Y. Aksyutina and H. {Álvarez-Pol} and L. Atar and T. Aumann and S. {Beceiro-Novo} and C. A. Bertulani and K. Boretzky and M. J. G. Borge and C. Caesar and M. Chartier and A. Chatillon and L. V. Chulkov and D. {Cortina-Gil} and P. {Díaz Fernández} and H. Emling and O. Ershova and L. M. Fraile and H. O. U. Fynbo and D. Galaviz and H. Geissel and M. Heil and M. Heine and D. H. H. Hoffmann and M. Holl and H.T. Johansson and B. Jonson and C. Karagiannis and O. A. Kiselev and J. V. Kratz and R. Kulessa and N. Kurz and C. Langer and M. Lantz and T. {Le Bleis} and R. Lemmon and Y. A. Litvinov and B. Löher and K. Mahata and J. {Marganiec-Gal} and C. Müntz and T. Nilsson and C. Nociforo and W. Ott and V. Panin and S. Paschalis and A. Perea and R. Plag and R. Reifarth and A. Richter and K. Riisager and C. {Rodriguez-Tajes} and D. Rossi and D. Savran and H. Scheit and G. Schrieder and P. Schrock and H. Simon and J. Stroth and K. Sümmerer and O. Tengblad and H. Weick and C. Wimmer",
   title        = "{Unveiling the two-proton halo character of $^{17}$Ne: Exclusive measurement of quasi-free proton-knockout reactions}",
   journal      = "Phys.\ Lett.\ B",
   volume       = "827",
   pages        = "136957",
   year         = "2022",
   url          = "https://doi.org/10.1016/j.physletb.2022.136957"
}

@ARTICLE{Zha25,
   author       = "K. Y. Zhang and X. X. Lu",
   title        = "{Microscopic description of the proton halo in $^{12}$N}",
   journal      = "Phys.\ Lett.\ B",
   volume       = "871",
   pages        = "139989",
   year         = "2025",
   url          = "https://doi.org/10.1016/j.physletb.2025.139989"
}

@ARTICLE{Sme99,
   author       = "M. H. Smedberg and T. Baumann and T. Aumann and L. Axelsson and U. Bergmann and M. J. G. Borge and D. {Cortina-Gil} and L. M. Fraile and H. Geissel and L. Grigorenko and M. Hellström and M. Ivanov and N. Iwasa and R. Janik and B. Jonson and H. Lenske and K. Markenroth and G. Münzenberg and T. Nilsson and A. Richter and K. Riisager and C. Scheidenberger and G. Schrieder and W. Schwab and H. Simon and B. Sitar and P. Strmen and K. Sümmerer and M. Winkler and M. V. Zhukov",
   title        = "{New results on the halo structure of $^{8}$B}",
   journal      = "Phys.\ Lett.\ B",
   volume       = "452",
   pages        = "1",
   year         = "1999",
   url          = "https://doi.org/10.1016/S0370-2693(99)00245-2"
}

@ARTICLE{Tan96,
   author       = "I. Tanihata",
   title        = "{Neutron halo nuclei}",
   journal      = "J.\ Phys.\ G",
   volume       = "22",
   pages        = "157",
   year         = "1996",
   url          = "https://doi.org/10.1088/0954-3899/22/2/004"
}

@ARTICLE{Jon04,
   author       = "B. Jonson",
   title        = "{Light dripline nuclei}",
   journal      = "Phys.\ Rep.",
   volume       = "389",
   pages        = "1",
   year         = "2004",
   url          = "https://doi.org/10.1016/j.physrep.2003.07.004"
}

@ARTICLE{Sun25,
   author       = "L. J. Sun and J. Dopfer and A. Adams and C. Wrede and A. Banerjee and B. A. Brown and J. Chen and E. A. M. Jensen and R. Mahajan and T. Rauscher and C. Sumithrarachchi and L. E. Weghorn and D. Weisshaar and T. Wheeler",
   title        = "{Extension of the particle x-ray coincidence technique: The lifetimes and branching ratios apparatus}",
   journal      = "Phys.\ Rev.\ C",
   volume       = "111",
   pages        = "055806",
   year         = "2025",
   url          = "https://doi.org/10.1103/PhysRevC.111.055806"
}

\end{document}